# Wave nature and metastability of emergent crystals in chiral magnets

Yangfan Hu[1]

Topological spin textures emerging in magnetic materials usually appear in crystalline states. A long-standing dilemma is whether we should understand these emergent crystals as gathering "particles" or coupling waves, the answer of which affects almost every aspect of our understanding on the subject. Here we prove that 2-D emergent crystals with long-range order in helimagnets, such as skyrmion crystals and dipole skyrmion crystals, have a wave nature. We systematically study their equilibrium properties, metastability, and phase transition path when unstable. We show that the robustness of a skyrmion crystal derives from its metastability, and that its phase transition dynamics at low (high) magnetic field is mediated by a soft mode which breaks (maintains) its hexagonal symmetry. Different from ordinary crystals which are formed by. and breaks into atoms, emergent crystals have a new formation (destruction) mechanism: they appear from (turn to) "single-Q" spin-density-wave states through nonlinear mode-mode interactions.

[1] Sino-French Institute of Nuclear Engineering and Technology, Sun Yat-sen University, 519082 Zhuhai, China. Correspondence and requests for materials should be addressed to Y.H. (email: huyf3@mail.sysu.edu.cn)

Magnetic skyrmions[1–3] are topological spin textures emerging in magnetic materials due to non-collinear interactions permitted by inversion symmetry breaking. They are attractive due to their high mobility to electric current[4,5], tunability to material size[6,7], and sensitivity to various kinds of field[8–12]. Besides ordinary skyrmions, other magnetic "emergent particles" have also been discovered in recent years: anti-skyrmions[13] are found in non-centrosymmetric ferromagnets (antiferromagnets) with $D_{2d}$ and $S_4$ symmetry, biskyrmions[14] are stabilized by demagnetization effects in common magnetic materials, and dipole-skyrmions[15,16] are found in magnetic multilayers with dipole interactions. Instead of isolated individuals, these emergent particles generally appear in crystalline states with long-range order[17] in most bulk materials[1,10,18] and layered structures[14–16,19–25] where they are detected. A fundamental question under debate is: should we understand emergent crystals as gathering particles[26–29] or coupling waves[1,17,18]? Since the local symmetry permitted by the field configuration of an emergent particle is usually incompatible with the global symmetry permitted by its emergent crystalline state, the answer to this question is vital for us to understand the basic properties of emergent crystals and their relation with the composing emergent particles.

In this work, we prove based on strict variational analysis that 2-D emergent crystals appearing in helimagnets with long-range order have a wave nature. Based on a wave description of four types of emergent crystals, we not only obtain their equilibrium properties, but also successfully reproduce their phase transition dynamics at both low and high magnetic fields. The phase transition behavior of skyrmion crystal (SkX) at low magnetic field reveals a novel formation (destruction) mechanism of emergent crystals, such that they emerge from (turn to) single-Q spin-density-wave states through non linear mode–mode interactions. This mechanism is not seen for ordinary crystals.

## Results

**Variational analysis of emergent crystals in helimagnets**. We perform our analysis based on a rescaled free energy density functional[1,27,30,31] for cubic helimagnets (see the Methods section for details)

$$\tilde{w}(\mathbf{m}) = \sum_{i=1}^{3}\left(\frac{\partial \mathbf{m}}{\partial r_i}\right)^2 + 2\mathbf{m}\cdot(\nabla\times\mathbf{m}) - 2\mathbf{b}\cdot\mathbf{m} + t\mathbf{m}^2 + \mathbf{m}^4, \quad (1)$$

where $\mathbf{m}$, $t$, and $\mathbf{b}$ denote respectively the rescaled magnetization vector, the rescaled temperature and the rescaled magnetic field. An emergent crystal appears as a periodic solution of $\mathbf{m}(r_i)$ that minimizes the averaged free energy density $\bar{w}(\mathbf{m}) = \frac{1}{V}\int \tilde{w}(\mathbf{m})dV$. At first glance, it is straightforward to expand $\mathbf{m}(r_i)$ in a Fourier series if we presuppose the existence of a periodic magnetization structure. However, when a function has discontinuous first order partial derivatives at certain points, the "Gibbs phenomenon"[32] is induced which makes it not suitable for Fourier analysis (see a detailed Fourier analysis of the circular cell approximation (CCA) [26] in Supplementary Note 1 and 2, where one should keep in mind that the CCA configuration keeps simultaneously the local axial symmetry of an isolated skyrmion and the global hexagonal symmetry of the lattice).

In this case, a proper question to ask is: to be a solution that minimizes $\bar{w}(\mathbf{m})$, can $\mathbf{m}$ have discontinuous first order partial derivatives at certain boundary in the unit cell? Concerning 2-D emergent particles only, the shape of this boundary should be a circle so that inside the circle the axial symmetry of $\mathbf{m}$ is maintained due to the axial symmetry of Eq. (1), and outside the circle the axial symmetry of $\mathbf{m}$ is destroyed due to long-range order of the emergent crystal. We construct the following field configuration in Fig. 1: in a hexagonal unit cell, $\mathbf{m}$ is axially symmetric for $\rho \le r_0$ (in the orange region), and is not axial symmetric for $\rho > r_0$ (in the blue region). The integrability condition of $\bar{w}(\mathbf{m})$ requires that $\mathbf{m}$ is continuous everywhere, and in both regions $0 < \rho < r_0$ and $\rho > r_0$, the first order derivatives of $\mathbf{m}$ with respect to the spatial variables are continuous. It is convenient to describe $\mathbf{m}$ in spherical coordinates by $(m, \theta, \psi)$ and to describe the space in cylindrical coordinates $(\rho, \varphi, z)$. At $\rho = 0$, local axial symmetry requires $\sin\theta = 0$, while $\psi_\rho = \frac{\partial\psi}{\partial\rho}$ does not need to be continuous. Since $\mathbf{m}$ is continuous on the interface $\rho = r_0$, $m_\varphi$, $\theta_\varphi$, and $\psi_\varphi$ are all continuous on the interface $\rho = r_0$. Such a field configuration of $\mathbf{m}$ keeps simultaneously the local axial symmetry in $\rho \le r_0$ and the global hexagonal symmetry. Yet, for such a field configuration of $\mathbf{m}$ to be the solution of the variational problem, additional conditions are required. Applying the Haar's theorem[32] in an arbitrary sector region $K$ depicted in Fig. 1 to the variational problem that minimizes $\bar{w}(\mathbf{m})$, we find that the solution function $m = m(\rho, \varphi)$, $\theta = \theta(\rho, \varphi)$, and $\psi = \psi(\rho, \varphi)$ must satisfy

$$\int_\alpha^\beta\int_{r_1}^{r_2}\rho\tilde{w}_\lambda d\rho d\varphi = \int_\alpha^\beta\left[r_2\left(\tilde{w}_{\lambda_\rho}\right)_{\rho=r_2} - r_1\left(\tilde{w}_{\lambda_\rho}\right)_{\rho=r_1}\right]d\varphi + \int_{r_1}^{r_2}\left[\rho\left(\tilde{w}_{\lambda_\rho}\right)_{\varphi=\beta} - \rho\left(\tilde{w}_{\lambda_\rho}\right)_{\varphi=\alpha}\right]d\rho, \quad (2)$$

where $\lambda$ is to be replaced by $m$, $\theta$, or $\psi$. Let $r_1 = r_0 - \varepsilon$, $r_2 = r_0 + \varepsilon$, and substituting Eq. (1) into Eq. (2), for $\varepsilon \to 0$ we have after manipulation

$$\lim_{\varepsilon\to 0}\left(m_\rho\right)_{\rho=r_0+\varepsilon} = \lim_{\varepsilon\to 0}\left(m_\rho\right)_{\rho=r_0-\varepsilon}, \quad (3)$$

$$\lim_{\varepsilon\to 0}\left(\theta_\rho\right)_{\rho=r_0+\varepsilon} = \lim_{\varepsilon\to 0}\left(\theta_\rho\right)_{\rho=r_0-\varepsilon}, \quad (4)$$

$$\lim_{\varepsilon\to 0}\left(\sin^2\theta\psi_\rho\right)_{\rho=r_0+\varepsilon} = \lim_{\varepsilon\to 0}\left(\sin^2\theta\psi_\rho\right)_{\rho=r_0-\varepsilon}, \quad (5)$$

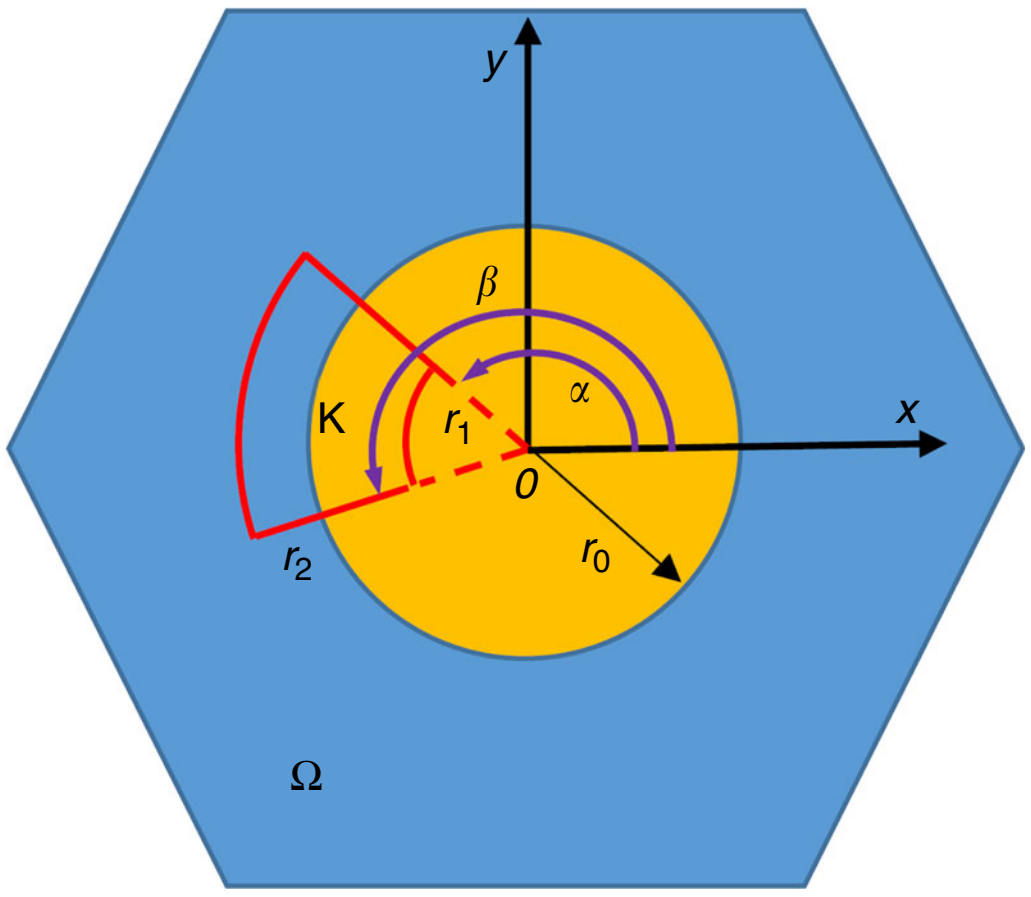

**Fig. 1** Illustration of a hexagonal unit cell Ω. In the orange circular region bounded by $\rho = r_0$, the field configuration of $\mathbf{m}$ is axially symmetric, while in the rest of the cell (blue region), the axial symmetry is broken. We construct a sector region $K$ bounded by $\rho = r_1$, $\rho = r_2$, $\varphi = \alpha$, and $\varphi = \beta$

**Table 1 Information about the Fourier representation for hexagonal lattice**

| $i$ | 1 | 2 | 3 | 4 | 5 | 6 | 7 | 8 |
|---|---|---|---|---|---|---|---|---|
| $n_i$ | 6 | 6 | 6 | 12 | 6 | 6 | 12 | 6 |
| $s_i$ | 1 | $\sqrt{3}$ | 2 | $\sqrt{7}$ | 3 | $2\sqrt{3}$ | $\sqrt{13}$ | 4 |
| $\mathbf{q}_{i1}$ | (1,0) | (0,$\sqrt{3}$) | (2,0) | (2,$\sqrt{3}$) | (3,0) | (0,2$\sqrt{3}$) | (1,2$\sqrt{3}$) | (4,0) |

where we have used the fact that $\alpha$ and $\beta$ are arbitrary. Equations (3) and (4) show that $m_\rho$ and $\theta_\rho$ are continuous on the interface $\rho = r_0$. Things are more complicated for Eq. (5), where we have at the interface $\rho = r_0$ either $\sin\theta = 0$ or $\psi_\rho$ to be continuous. Nevertheless, we find after derivation that in both cases ((1) $\sin\theta = 0$ and $\psi_\rho$ incontinuous; (2) $\psi_\rho$ continuous), $\nabla\mathbf{m}$ and $\nabla\times\mathbf{m}$ are continuous at $\rho = 0$ and $\rho = r_0$, which makes them continuous everywhere. For a continuous periodic vector function $\mathbf{v}$, its Fourier series expansion $\mathbf{v}_{\mathrm{F}n}$ is uniformly convergent to $\mathbf{v}$ as $n\to\infty$. As a result, the solution function $\mathbf{m}_{\min}$ of the variational problem, as well as $\nabla\mathbf{m}_{\min}$ and $\nabla\times\mathbf{m}_{\min}$ can all be expressed as Fourier series. Since Fourier series expansion and differentiation commute[33], they are uniformly approached by $\mathbf{m}_{\mathrm{F}n}$, $\nabla\mathbf{m}_{m\mathrm{F}n}$, and $\nabla\times\mathbf{m}_{m\mathrm{F}n}$, where $\mathbf{m}_{m\mathrm{F}n}$ is the Fourier series expansion of $\mathbf{m}_{\min}$. The minimized rescaled free energy density $\tilde{w}(\mathbf{m}_{\min})$, as a continuous function of $\mathbf{m}_{\min}$, $\nabla\mathbf{m}_{\min}$, and $\nabla\times\mathbf{m}_{\min}$, is therefore uniformly approached by $\tilde{w}(\mathbf{m}_{m\mathrm{F}n})$. Thus the Fourier representation is the exact form of solution to describe any emergent crystals in chiral magnets.

**Fourier representation of an emergent crystalline state**. The proof given above leads naturally to the following two points: (a) crystallization of isolated skyrmions into SkX is a phase transition, characterized by breaking of the local axial symmetry of an isolated skyrmion. (b) In general, the order parameter of any emergent crystals (e.g., $\mathbf{m}$ of SkX) can be described by the $n$th order Fourier representation as

$$\mathbf{m}_{\mathrm{F}n} = \mathbf{m}_0 + \sum_{i=1}^{n}\sum_{j=1}^{n_i}\mathbf{m}_{\mathbf{q}_{ij}} e^{\mathrm{i}\mathbf{q}_{ij}\cdot\mathbf{r}}, \quad (6)$$

where $|\mathbf{q}_{i1}| = |\mathbf{q}_{i2}| = |\mathbf{q}_{i3}| = \ldots = |\mathbf{q}_{in_i}| = s_i q$, $|\mathbf{q}_{1j}| < |\mathbf{q}_{2j}| < |\mathbf{q}_{3j}| < \ldots < |\mathbf{q}_{nj}|$, and $n_i$ denotes the number of reciprocal vectors whose modulus equal to $s_i q$. One should notice that such a description is generally applicable to any $n$-dimensional emergent crystals ($n = 1, 2, 3$). Some specific information about the Fourier representation of 2-D hexagonal lattices is listed inTable 1. We describe four types of emergent crystals here using Eq. (6), including the dipole skyrmion crystal (DiskX, 2-D emergent crystal, Fig. 2a, b), the SkX (2D emergent crystal, Fig. 2d, e), the in-plane single-Q phase (IPSQ phase, 1-D emergent crystal, Fig. 2h), and the ferromagnetic phase (trivial solution, Fig. 2i). The SkX shown in Fig. 2e and the DiskX shown in Fig. 2a, b are distinguished by their symmetry of field configuration: under the spatial inversion transformation $\mathbf{r}\to-\mathbf{r}$, we have for the SkX phase $m_1\to-m_1$, $m_2\to-m_2$, $m_3\to m_3$, and for the DiskX phase $m_1\to m_1$, $m_2\to m_2$, and $m_3\to-m_3$. Within the Fourier representation, all emergent crystals are described by the same set of independent variables $\mathbf{m}_0$, $\mathbf{m}_{\mathbf{q}_{ij}}$, and $q$, with different values.

Specific attention should be paid on the modulus of magnetization given in Eq. (6). Within a truncated Fourier representation, we find that the modulus of magnetization can never be a constant in space if $\mathbf{m}_0$ and $\mathbf{m}_{\mathbf{q}_{ij}}$ are all non-zero vectors. For example, the term $\sum_{j=1}^{n_i}\mathbf{m}_0\cdot\mathbf{m}_{\mathbf{q}_{1j}}e^{\mathrm{i}\mathbf{q}_{1j}\cdot\mathbf{r}}$ appearing in the expansion of $|\mathbf{m}_{\mathrm{F}n}|^2$ can never be canceled by increasing the Fourier expansion order. Further analysis tells us that $|\mathbf{m}_{\mathrm{F}n}|^2$ approaches a constant only when all $\mathbf{m}_{\mathbf{q}_{ij}}$ approaches zero (ferromagnetic phase). This modulations of the magnetization modulus is found to be important to understand the existence of precursor states near the ordering temperature discovered in previous studies within the functional given in Eq. (1)[34,35].

**Equilibrium properties of emergent crystals**. The Fourier representation provides an analytical framework to study the equilibrium properties of any emergent crystal. For a specific type of emergent crystal, one calculates the partial derivative of $\bar{w}(\mathbf{m})$ with respect to $\mathbf{m}_0$, $\mathbf{m}_{\mathbf{q}_{ij}}$, and $q$, which yields a set of high order algebraic equations. For the three types of non-trivial emergent crystals, we calculate the variation of $\bar{w}(\mathbf{m})$ with $b$ at $t = 0.5$ and compare it with that for the conical phase in Fig. 3a. Within mean-field approximation, the conical phase always has the lowest averaged free energy density, while DiskX always has the highest averaged free energy density. For all emergent crystals, the averaged free energy density is always slightly decreased for an increase of the Fourier representation order, in which case 1st order Fourier representation is found to be an effective approximation. On the other hand, the variation of lattice constant of emergent crystals with $b$ is found to be very sensitive to Fourier representation order (Fig. 3b). To be more specific, the $\mathbf{m}^4$ term in Eq. (1) permits non-zero mode–mode interactions, such as that described by $\mathbf{m}_0\cdot\mathbf{m}_{\mathbf{q}_{3k}}\mathbf{m}_{\mathbf{q}_{1j}}\cdot\mathbf{m}_{\mathbf{q}_{1j}}$, which is negative for appropriate choice of $\mathbf{m}_{\mathbf{q}_{3k}}$. Meanwhile, non-zero $\mathbf{m}_{\mathbf{q}_{3k}}$ reduces the equilibrium value of $q$ through the terms $\sum_{i=1}^{3}\left(\frac{\partial\mathbf{m}}{\partial r_i}\right)^2 + 2\mathbf{m}\cdot(\nabla\times\mathbf{m})$ in Eq. (1). As the reciprocal of $q$, the lattice constant generally increases with any $\left|\mathbf{m}_{\mathbf{q}_{ij}}\right|$, $(i\geq 2)$. For $n\geq 3$, the variation of lattice constant of SkX with magnetic field obtained in experiment[33] is well explained within the $n$th order Fourier representation of the SkX.

The equilibrium properties of the SkX phase is determined through numerically solving a set of algebraic equations, where the equations are derived analytically based on the Fourier representation of SkX. Nevertheless, this Fourier-based method is fundamentally different from pure numerical methods for free energy minimization such as the finite difference method and the Monte Carlo simulation method, since it provides the analytical expression of the magnetization. Due to the lack of an analytical expression in pure numerical methods, all information of SkX has to be extracted vaguely from the numerical distribution of magnetization, which is sometimes unreliable. For instance, the core region of a skyrmion in the SkX phase was thought to possess axial symmetry according to finite difference calculation[35]. From the Fourier representation of SkX, we learn that this is incorrect once SkX maintains strict long-range order, because a Fourier series destroys axial symmetry regardless of the expansion order. Moreover, similar to pure numerical methods, the precision of the Fourier-based method can be improved by increasing the Fourier expansion order, and when achieving the same level of precision, we find that the Fourier-based method is much more efficient than the Monte Carlo program we developed (See Supplementary Note 3 for more details).

**Metastability of emergent crystals**. One exotic property of SkX is that it is sometimes found to be stable at conditions where it is not supposed to be[18]. We explain this robustness of SkX by its metastability: SkX corresponds to a local minimum of the free energy surrounded by high barriers, and the thermal fluctuation is not strong enough for the system to escape from this trap. In this sense, the critical condition when emergent crystals losses

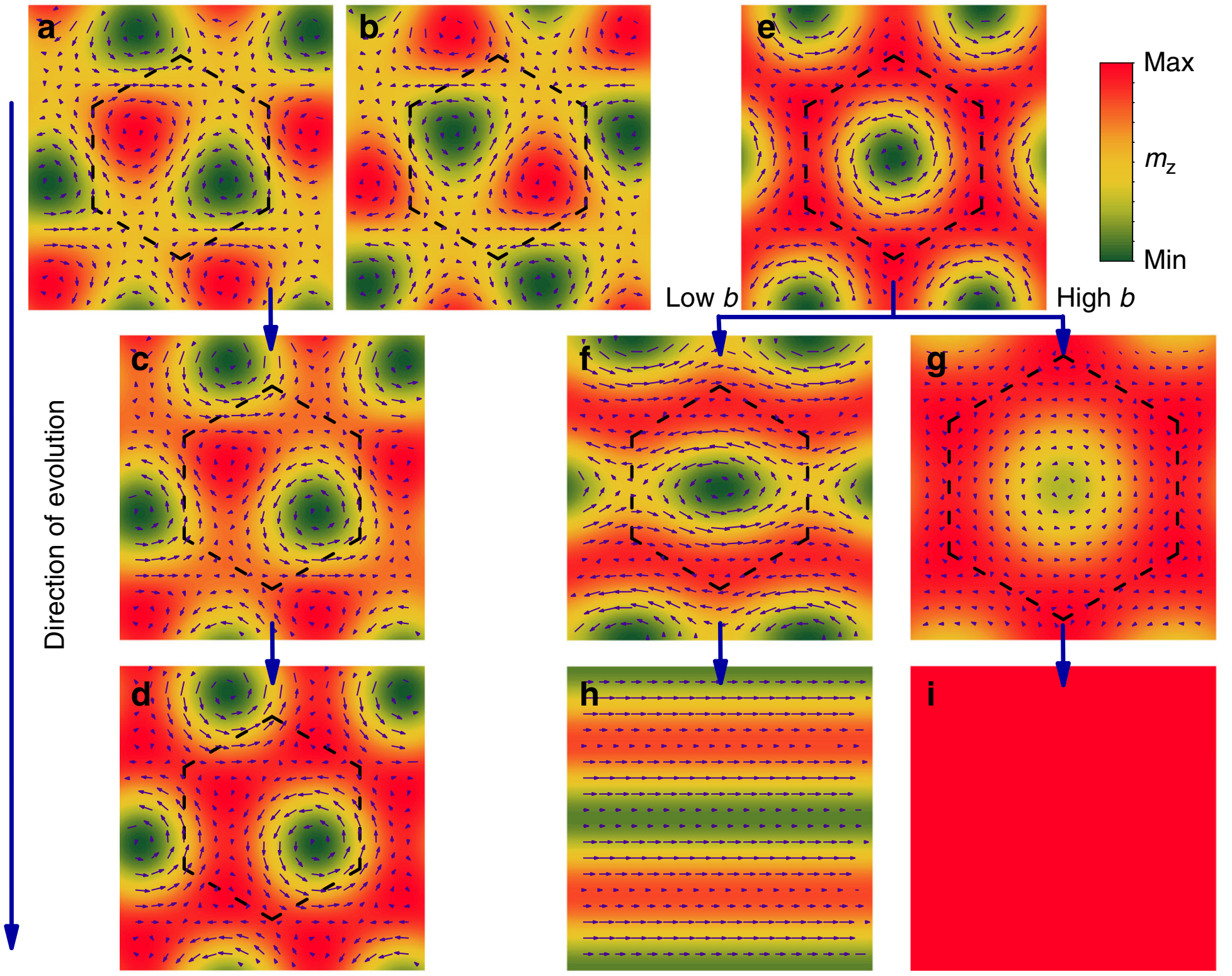

**Fig. 2** Field configurations and phase transition paths of emergent crystals. In all figures, the hexagon plotted with black dash-line represents the Wigner–Seitz cell of the emergent crystal concerned. The vectors illustrate the distribution of the in-plane magnetization components with length proportional to the magnitude, while the colored density plot illustrates the distribution of the out-of-plane magnetization component. **a** The dipole skyrmion crystal (DiskX) with a characteristic vector $\mathbf{V}_{Cb1} = \frac{\sqrt{3}}{3}[0\,0\,0\,1\,1\,1]^T$ will always evolve spontaneously to **d**, the skyrmion crystal (SkX) with a characteristic vector $\mathbf{V}_{C2} = [0.288\text{-}0.288\ -0.288\ 0.5\ 0.5\ 0.5]^T$ passing by **c**, an intermediate state with a characteristic vector $\mathbf{V}_{Cb1} + v_b\mathbf{V}_{Rb}$, where $\mathbf{V}_{Rb} = \frac{\sqrt{3}}{3}[1-1-1\,0\,0\,0]^T$ and $v_b = 0.275$. **b** An alternative DiskX with a characteristic vector $\mathbf{V}_{Cb2} = \frac{\sqrt{3}}{3}[0\,0\,0-1-1-1]^T$. **e** The SkX with a characteristic vector $\mathbf{V}_{C1} = \frac{\sqrt{3}}{3}[1\,1\,1\,0\,0\,0]^T$ will evolve spontaneously to **h**, the in-plane single-Q (IPSQ) phase when the magnetic field is smaller than $b_{csl}$, passing by **f**, an intermediate state with a characteristic vector $\mathbf{V}_{C1} + v_1\mathbf{V}_{R1}$, where $\mathbf{V}_{R1} = \frac{\sqrt{6}}{6}[-1\,2-1\,0\,0\,0]^T$, and $v_1 = 0.566$. When the magnetic field is above $b_{csh}$, **e** the SkX will evolve spontaneously to **i**, the ferromagnetic phase passing by **g**, an intermediate state with the same characteristic vector as **e**. The SkX–IPSQ transition (**e**, **f**, **h**) maintains the lattice constant while the SkX-ferromagnetic transition (**e**, **g**, **i**) is accompanied by an expansion

their metastability provides an upper bound concerning their robustness. In the Methods section, we establish the method of soft-mode analysis to determine this critical condition for emergent crystals. Applying the method to SkX at the condition $t = 0.7$, $b > 0$, we find that the metastability is dominated by two different modes at low and high magnetic field, which turn to soft modes at two critical magnetic fields $b_{csl}$ and $b_{csh}$ (Fig. 4a, b). The variation of $b_{csl}$ and $b_{csh}$ with $t$ determines the region of metastability of SkX in the $t$–$b$ phase diagram (Fig. 4c).

Similar analysis to the IPSQ phase and ferromagnetic phase yields two different critical magnetic fields $b_{ci}$ and $b_{cf}$. The discrepancy between $b_{ci}$ and $b_{csl}$ ($b_{cf}$ and $b_{csh}$) provides a region in the $t$–$b$ phase diagram (shadowed area in Fig. 4c) where both the IPSQ phase and SkX (the ferromagnet phase and SkX) are metastable, so no phase transition between the two states will spontaneously occur, and a two-phase state may be observed. Such a two-phase state of the IPSQ phase and SkX (the ferromagnet phase and SkX) is observed in TEM experiments of $Fe_{0.5}Co_{0.5}Si$[19] and FeGe[21] thin films, while a two-phase state of the DiskX and the IPSQ phase is also observed in $La_{2-2x}Sr_{1+2x}Mn_2O_7$[14] and $(Mn_{1-x}Ni_x)_{65}Ga_{35}$[25]. In a two-phase state, SkX and the IPSQ phase occupy different parts of the space, forming domain-like structures. Near the domain walls, intermediate state such as those depicted in Fig. 2f will appear locally.

It is shown in Fig. 3b that the most stable phase always corresponds to the conical phase at the condition $t = 0.5, 0 < b < 0.25$. We find that the conical phase is approached through a spin reorientation transition from the IPSQ phase. To account for this transition, one has to introduce the wave-vector-field angle into the expression of single-Q magnetization to describe the so-called general conical phase[31]. One should notice that although the conical phase corresponds to the state with global minimum of free energy at low magnetic field, the SkX phase always evolves to the IPSQ phase first according to the soft-mode analysis. This is because in the free energy landscape, the IPSQ phase is "closer" to the SkX phase than the conical phase in the sense that the first two states share the same form of Fourier representation.

**Phase transition dynamics of skyrmion crystal**. Another interesting phenomenon about SkX is the distinct phase transition behavior it shows at low and high magnetic field. TEM experiments of $Fe_{0.5}Co_{0.5}Si$[19] and FeGe[21] thin films both show that SkX becomes elongated in a certain direction at enough low magnetic field and finally evolves to the IPSQ phase, while at high enough magnetic field it evolves to the ferromagnetic phase without a distortion of the skyrmions. We reproduce the two phase transition paths of SkX explicitly in Fig. 2. The two different transition paths can be understood by investigating the characteristic vectors $\mathbf{V}_C$ of the corresponding soft modes. According to its definition, a phase transition of SkX maintains the hexagonal symmetry if $\mathbf{V}_C$ of the corresponding soft mode satisfies $(V_C)_1 =$

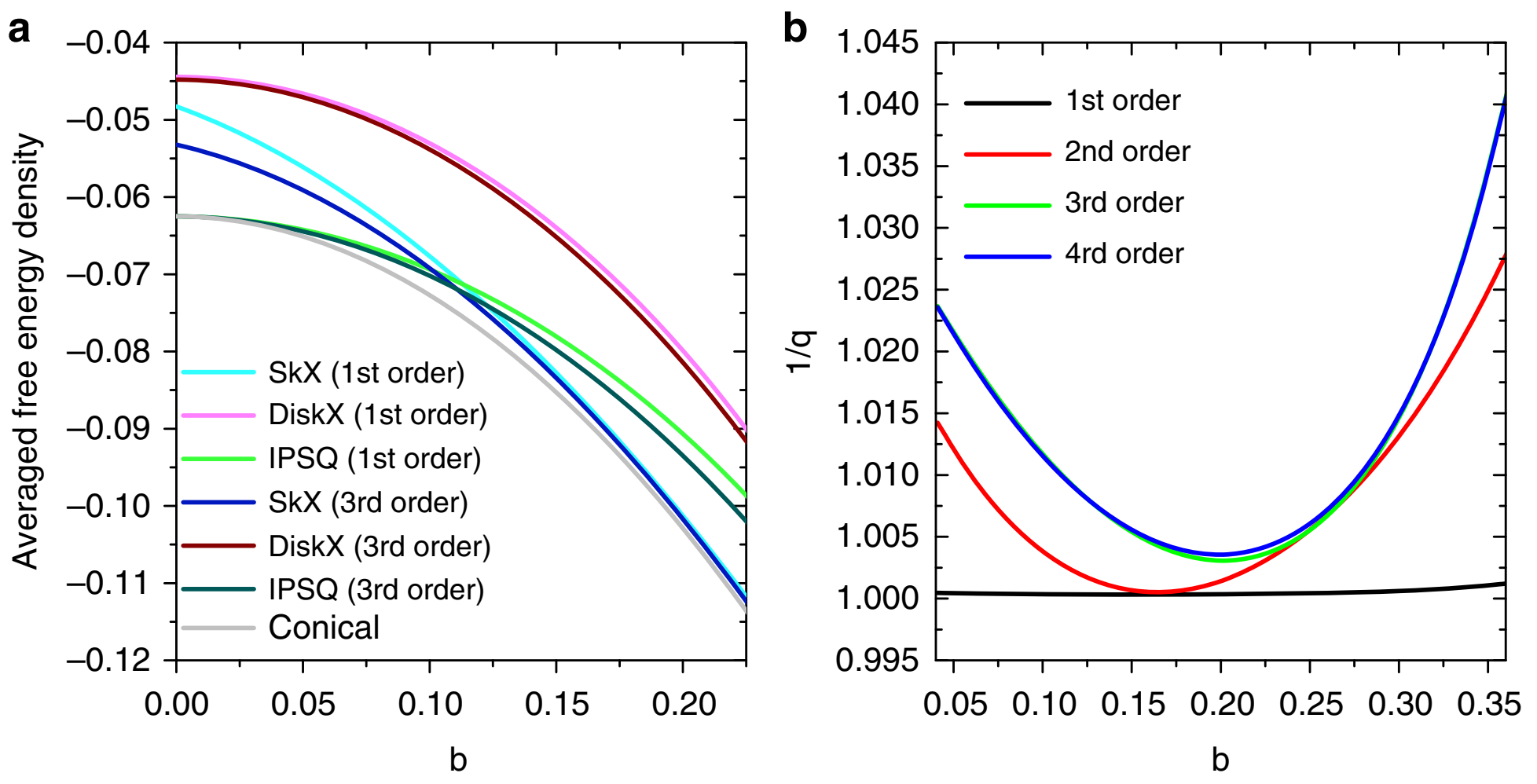

**Fig. 3** Equilibrium properties of different magnetic phases. All calculations are done at $t = 0.5$, $b > 0$. **a** Variation of $\bar{w}$ with the rescaled magnetic field $b$ for several magnetic phases. **b** The non-dimensionalized lattice constant of skyrmion crystal (SkX) versus the rescaled magnetic field $b$ calculated within different orders of Fourier representation

$(V_C)_2 = (V_C)_3$ and $(V_C)_4 = (V_C)_5 = (V_C)_6$, which explains the different phase transition paths of SkX at low and high magnetic fields shown in Fig. 2. Specifically, the mechanism for the SkX–IPSQ transition lies in the non-linear mode–mode interaction. Within the 1st order Fourier representation, the triple-Q SkX induces two more types of coupling compared with the IPSQ phase due to the $\mathbf{m}^4$ term in the free energy density, one represented by $\mathbf{m}_{q_{11}}^2 \mathbf{m}_{q_{12}}^2$, and the other represented by $\mathbf{m}_0 \cdot \mathbf{m}_{q_{13}} \mathbf{m}_{q_{11}} \cdot \mathbf{m}_{q_{12}}$. The first type of interaction always increases the free energy density of SkX compared with that of the IPSQ phase, while the second type of interaction can effectively reduce the free energy density of SkX by appropriately choosing the direction between $\mathbf{m}_0$ and $\mathbf{m}_{q_{11}}, \mathbf{m}_{q_{12}}, \mathbf{m}_{q_{13}}$. Since $|\mathbf{m}_0| \to 0$ when $b \to 0$, the second type of interaction has a negligible effect when the magnetic field is low enough, for which the IPSQ phase is more stable than SkX. As $b$ increases, the second type of interaction becomes more significant, and at certain point SkX becomes more stable. Since interaction between waves occurs globally, the stability of SkX at low magnetic field is found to be irrelevant to the local topological protection of an isolated skyrmion. Moreover, it implies that emergent crystals composed of emergent particles that are unstable alone may appear under appropriate conditions, which is supported by the phase transition from a triangular SkX to a square SkX observed in a Co–Z–Mn alloy[18].

Applying the method of soft-mode analysis to DiskX, we find that DiskX is always unstable in chiral magnets (Fig. 4) and spontaneously evolve to SkX with a phase transition path shown in Fig. 2a, c, d. The triangle chiral whirl with magnetization pointing downwards (green whirl in Fig. 2a) will slightly enlarge and finally become a skyrmion with hexagonal shape in SkX.

**Wave-particle duality for emergent crystals**. The analysis above shows that the wave description of an emergent crystal is not only mathematically correct, but also physically indispensable. The phase transition of SkX to the IPSQ state at low magnetic field, mediated by softening of macroscopic spin-density-waves, cannot be understood by merely studying the localized skyrmion–skyrmion interaction. Moreover, the wave description of an emergent crystal does not deny the particle nature of its localized composing units. The phase transition of SkX to the ferromagnetic phase at high magnetic field is mediated by a diverging expansion of the emergent crystal lattice which maintains the crystalline symmetry (Fig. 2g). This result coincides with previous analysis[34,35], indicating a crossover phenomenon of the skyrmion–skyrmion interaction from mutual attractive to mutual repulsive. In this process, isolated skyrmion should appear as an intermediate state when the skyrmion–skyrmion interaction is weak enough. These two phase transitions show precisely the wave-particle duality of the SkX phase, which is also a fundamental difference between emergent crystals and atomic crystals. Unlike atomic crystals which are always condensates of atoms, emergent crystals can be formed either by closed packing of localized emergent particles, or by interacting macroscopic density waves with different wave vectors.

The proof of wave nature of emergent crytals relies on strict long-range order of the crystalline states. In real materials, existence of fluctuation, boundaries, defects, and microstructures all affects this precondition, for which an emergent particle in an emergent crystal may present particle properties locally. For instance, the L-shaped elongated skyrmions observed in supercooled $Co_8Zn_8Mn_4$ thin plates as a metastable phase[36] shows simultaneously a tendency towards the IPSQ phase, as well as the presence of local topological protection of the composing skyrmions. To deal with these effects, the present theoretical method needs to be developed. When the effect of thermal fluctuation is small enough to be regarded as a perturbation of the equilibrium state, the saddle-point method provides the first order approximation to account for the effect of fluctuation on the equilibrium properties[1,37]. Yet, when the long-range order is severely damaged, such as in the skyrmion glass state[38], it will be extremely hard to construct an effective method based on modification of the present theory. Similar discussion can be made on the effect of boundary and sample size. We know that as the diameter of a nanodisk of chiral magnets increases, the magnetic state evolves from an isolated skyrmion to skyrmion clusters and then to SkX[39]. This example shows the competition between boundary effects and bulk effects. In magnetic states where the boundary effects dominate (i.e., isolated skyrmion and skyrmion clusters), the magnetic state depends sensitively on the shape and conditions of the boundary, for which the wave description may not be a good starting point towards an effective theory.

## Discussion

We prove based on variational analysis that emergent crystals appearing in chiral magnets with long-range order has a wave

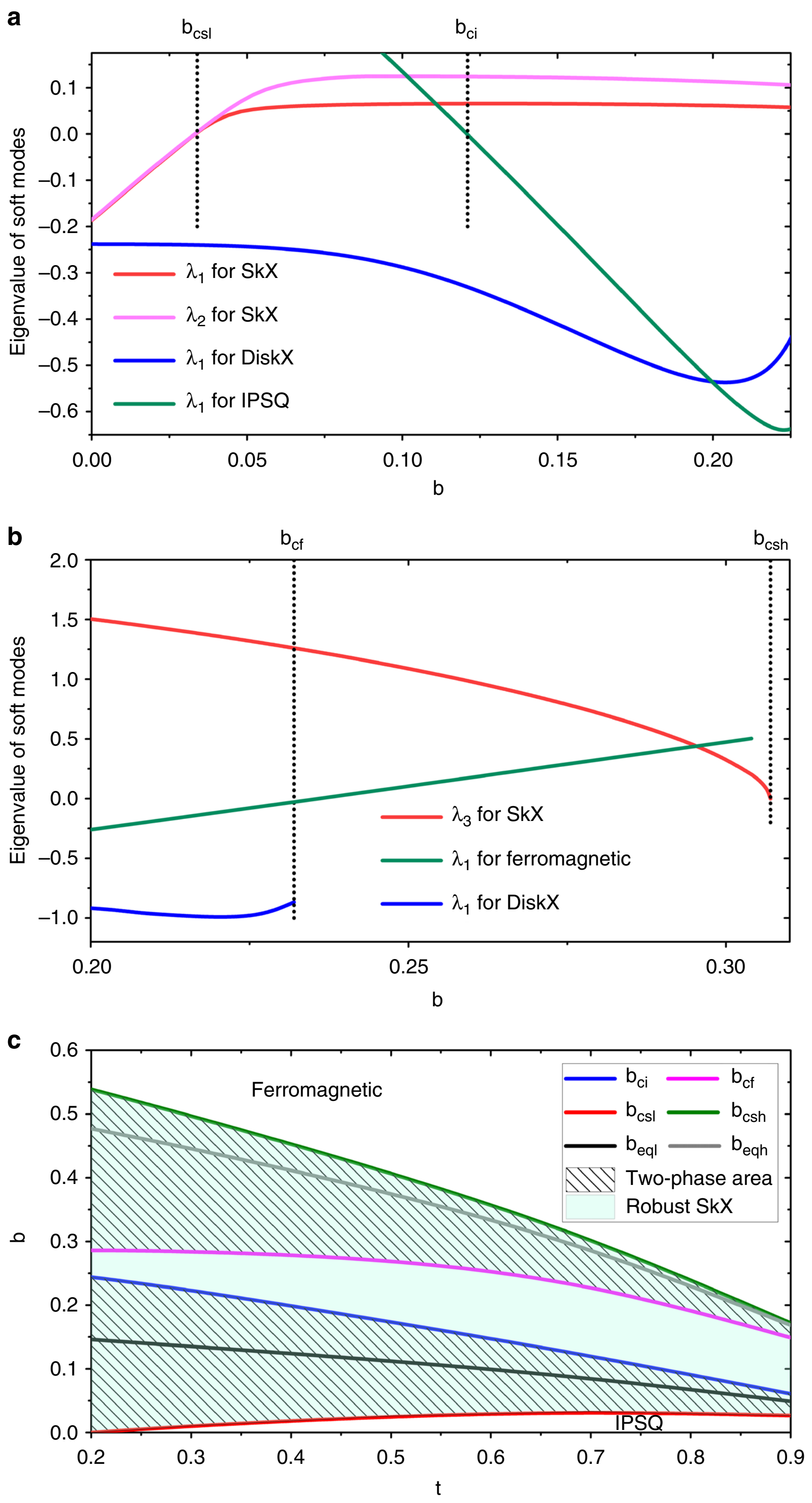

**Fig. 4** Results of soft-mode analysis for different emergent crystals. **a** Variation of the eigenvalues of the dominant soft modes with the rescaled magnet field $b$ for the skyrmion crystal (SkX), the dipole skyrmion crystal (DiskX), and the in-plane single-Q (IPSQ) phase calculated for low magnetic field at $t = 0.7$. $b_{\mathrm{csl}}$ ($b_{\mathrm{ci}}$) denotes the critical magnetic field for a spontaneous SkX–IPSQ (IPSQ–SkX) phase transition. **b** Variation of the eigenvalues of the dominant soft modes with $b$ for the SkX, the DiskX and the ferromagnetic phase calculated for high magnetic field at $t = 0.7$. $b_{\mathrm{csh}}$ ($b_{\mathrm{cf}}$) denotes the critical magnetic field for a spontaneous SkX-ferromagnetic (ferromagnetic-SkX) phase transition. **c** Phase diagram of metastabiliy for the SkX phase, where $b_{\mathrm{eql}}$ ($b_{\mathrm{eqh}}$) denotes the magnetic field at which the SkX phase and the IPSQ (ferromagnetic) phase corresponds to the same averaged free energy density. The shadowed area marks the region where a two-phase state may appear. The light green area marks the region where SkX may exist metastably

nature, such that their exact mathematical expression is a Fourier series. Then we construct a systematic approach upon this analytical Fourier framework to obtain the equilibrium properties, the metastabiliy region in the phase diagram, and the phase transition path of any types of emergent crystals. We apply the method to analyze four types of emergent crystals, including the SkX, DiskX, the IPSQ phase, and the ferromagnetic phase. For SkX, we successfully explain its phase transition behaviors, such that it transforms to the IPSQ phase at low magnetic field and to the ferromagnetic phase at high magnetic field. We show that emergent crystals can be formed by interaction between spin-density waves with different wave vectors. This new formation mechanism differs emergent crystals from ordinary atomic crystals.

## Methods

**Free energy functional and rescaling process**. We use the following free energy density functional to study magnetic skyrmions in cubic helimagnets

$$w(\mathbf{M}) = \sum_{i=1}^{3} A\left(\frac{\partial \mathbf{M}}{\partial x_i}\right)^2 + D\mathbf{M}\cdot(\nabla\times\mathbf{M}) - \mathbf{B}\cdot\mathbf{M} + \alpha(T-T_0)\mathbf{M}^2 + \beta\mathbf{M}^4, \quad (7)$$

where $\mathbf{M}$ denotes the magnetization vector, $\mathbf{B}$ denotes the external magnetic field, and $T$ denotes the temperature. The five terms on the rhs. of Eq. (7) denote respectively the exchange energy density, the Dzyaloshinskii–Moriya (DM) interaction, the Zeeman energy density, and the second and fourth order Landau expansion terms. Eq. (7) can be reduced to $\tilde{w}(\mathbf{m}) = \frac{\beta}{K^2}w(\mathbf{m})$ in Eq. (1) by rescaling the spatial variables as

$$\mathbf{r} = \frac{\mathbf{x}}{L_D},\ \mathbf{b} = \frac{\mathbf{B}}{B},\ \mathbf{m} = \frac{\mathbf{M}}{M_0},\ L_D = \frac{2A}{D},\ B = 2KM_0,\quad M_0 = \sqrt{\frac{K}{\beta}},\ K = \frac{D^2}{4A},\ t = \frac{\alpha(T-T_0)}{K}. \quad (8)$$

To derive Eqs. (3)–(5), Eq. (1) should be described in the spherical coordinates as

$$\begin{aligned}\tilde{w} = {} & m^2\left[\left(\theta_\rho\right)^2 + \left(\psi_\rho\right)^2\sin^2\theta + \frac{\left(\theta_\varphi\right)^2}{\rho^2} + \frac{\left(\psi_\varphi\right)^2\sin^2\theta}{\rho^2}\right] \\ & + \left(m_\rho\right)^2 + \frac{\left(m_\varphi\right)^2}{\rho^2} + \frac{2m^2}{\rho}\Big[-\cos(\psi-\varphi)\theta_\varphi \\ & + \sin\theta\cos\theta\cos(\psi-\varphi)\psi_\varphi\Big] + 2m^2\Big[\sin(\psi-\varphi)\theta_\rho \\ & + \sin\theta\cos\theta\cos(\psi-\varphi)\psi_\rho\Big] - 2bm\cos\theta + tm^2 + m^4.\end{aligned} \quad (9)$$

**Solving the equilibrium properties of an emergent crystal**. Substituting the $n$th order Fourier representation of emergent crystals into Eq. (1), after integration we have

$$\bar{w}(\mathbf{m}_{\mathrm{Fn}}) = -\frac{(t-1)^2}{4} - b^2 + (\mathbf{m}_0 - \mathbf{b})^2 + \bar{w}_{\mathrm{int}} + \bar{w}_{\mathrm{per}}, \quad (10)$$

where $\bar{w}(\mathbf{m}_{\mathrm{Fn}}) = \frac{1}{V}\int\tilde{w}(\mathbf{m}_{\mathrm{Fn}})\mathrm{d}V$, $\bar{w}_{\mathrm{int}} = \frac{1}{V}\int\left(\mathbf{m}_{\mathrm{Fn}}^2 + \frac{t-1}{2}\right)^2\mathrm{d}V$ includes the effect of non-harmonic mode–mode interaction, and $\bar{w}_{\mathrm{per}} = \sum_{i=1}^{n}\sum_{j=1}^{n_i}\left(\mathbf{m}^*_{\mathbf{q}ij}\right)^T\mathbf{A}_{ij}\mathbf{m}_{\mathbf{q}ij}$ includes the effect of all the gradient terms. Here $\mathbf{m}^*_{\mathbf{q}ij}$ denotes the conjugate of $\mathbf{m}_{\mathbf{q}ij}$, and

$$\mathbf{A}_{ij} = \begin{bmatrix} 1+s_i^2q^2 & 0 & 2\mathrm{i}q_{ijy} \\ 0 & 1+s_i^2q^2 & -2\mathrm{i}q_{ijx} \\ -2\mathrm{i}q_{ijy} & 2\mathrm{i}q_{ijx} & 1+s_i^2q^2 \end{bmatrix}, \quad (11)$$

where $\left|\mathbf{q}_{ij}\right| = s_iq = \sqrt{q_{ijx}^2 + q_{ijy}^2}$ and $\mathrm{i} = \sqrt{-1}$. The eigenvalues of $\mathbf{A}_{ij}$ are $(s_iq-1)^2$, $1+s_i^2q^2$, and $(s_iq+1)^2$, and the corresponding unit eigenvectors can be solved as $\mathbf{P}_{ij1} = \frac{1}{\sqrt{2}s_iq}\left[-\mathrm{i}q_{ijy}, \mathrm{i}q_{ijx}, s_iq\right]^T$, $\mathbf{P}_{ij2} = \frac{1}{s_iq}\left[q_{ijx}, q_{ijy}, 0\right]^T$, $\mathbf{P}_{ij3} = \frac{1}{\sqrt{2}s_iq}\left[\mathrm{i}q_{ijy}, -\mathrm{i}q_{ijx}, s_iq\right]^T$. Hence $\mathbf{m}_{\mathbf{q}ij}$ can be expanded as a linear combination of the three unit eigenvectors

$$\mathbf{m}_{\mathbf{q}ij} = c_{ij1}\mathbf{P}_{ij1} + c_{ij2}\mathbf{P}_{ij2} + c_{ij3}\mathbf{P}_{ij3}, \quad (12)$$

where $c_{ijk} = c_{ijk}^{\mathrm{re}} + \mathrm{i}c_{ijk}^{\mathrm{im}}$ and $c_{ijk}^{\mathrm{re}}$ and $c_{ijk}^{\mathrm{im}}$ are real variables to be determined. Using Eq. (12), $\bar{w}_{\mathrm{per}}$ can be recasted as $\sum_{i=1}^{n}\sum_{j=1}^{n_i}\left[\left|c_{ij1}\right|^2(s_iq-1)^2 + \left|c_{ij2}\right|^2\left(1+s_i^2q^2\right) + \left|c_{ij3}\right|^2(s_iq+1)^2\right]$, which is positive semidefinite. By solving $\bar{w}_{\mathrm{per}} = 0$, we obtain a unique solution $q = 1$, $|c_{ij2}| = |c_{ij3}| = 0$ for all $i$, $j$, and $|c_{ij1}| = 0$ for $i \neq 1$. This provides us with a valuable start-point to search for any possible emergent crystalline state in chiral magnets described by the free energy density functional given in Eq. (1). Taking 1st order Fourier representation for triangle lattice as an example, there are 22 independent variables to describe an emergent crystal, which can be written in a vector form as

$$\mathbf{V} = \begin{bmatrix} m_{01} & m_{02} & m_{03} & c_{111}^{\mathrm{re}} & c_{121}^{\mathrm{re}} & c_{131}^{\mathrm{re}} & c_{112}^{\mathrm{re}} & \cdots & c_{133}^{\mathrm{re}} & c_{111}^{\mathrm{im}} & c_{121}^{\mathrm{im}} & \ldots & c_{133}^{\mathrm{im}} & q \end{bmatrix} \quad (13)$$

Assume that $\mathbf{b} = \begin{bmatrix}0 & 0 & b\end{bmatrix}^T$, then the dominant independent variables can be written in a vector form with eight components as

$$\mathbf{V}_{\mathrm{D}} = \begin{bmatrix} m_{03} & c_{111}^{\mathrm{re}} & c_{121}^{\mathrm{re}} & c_{131}^{\mathrm{re}} & c_{111}^{\mathrm{im}} & c_{121}^{\mathrm{im}} & c_{131}^{\mathrm{im}} & q \end{bmatrix}^T. \quad (14)$$

For 2nd order Fourier representation, the number of components of $\mathbf{V}$ increases to 40, and for 3rd order Fourier representation it increases to 58, while the number of components of the dominant vector $\mathbf{V}_{\mathrm{D}}$ does not change with the order of Fourier representation. The equilibrium state of any emergent crystal is determined by solve

$$\frac{\partial \bar{w}}{\partial V_i} = 0,\ (i = 1, 2, \ldots), \quad (15)$$

which yields a set of $n$-element algebraic equations. In general, the number $n$ is too large (for the lowest order, $n = 22$) to derive an analytical solution of Eq. (15) directly. Numerically, an effective and efficient way is to start by solving $\frac{\partial \bar{w}}{\partial V_{\mathrm{D}i}} = 0, (i = 1, 2, \ldots, 8)$, while setting all other variables not included in $\mathbf{V}_{\mathrm{D}}$ as zero in the first place. Such a solution does not minimize $\bar{w}$, but is usually near by a local minimum of $\bar{w}$. Any local minimum of $\bar{w}$ is determined by a competition between $\bar{w}_{\mathrm{per}}$ and $\bar{w}_{\mathrm{int}}$. The solution of $\frac{\partial \bar{w}}{\partial V_{\mathrm{D}i}} = 0,\ (i = 1, 2, \ldots, 8)$ minimizes $\bar{w}_{\mathrm{per}}$ but not $\bar{w}_{\mathrm{int}}$. To be more specific, the $\mathbf{m}^4$ term in $\bar{w}_{\mathrm{int}}$ permits non-zero mode–mode interactions, such as that described by $\mathbf{m}_0 \cdot \mathbf{m}_{\mathbf{q}3k}\mathbf{m}_{\mathbf{q}1j} \cdot \mathbf{m}_{\mathbf{q}1j}$, which can effectively reduce the free energy by appropriately choosing $\mathbf{m}_{\mathbf{q}3k}$. Meanwhile, non-zero $\mathbf{m}_{\mathbf{q}3k}$ reduces the equilibrium value of $q$ through $\bar{w}_{\mathrm{per}}$, leading to a variation of the lattice constant for emergent crystals. Nevertheless, since non-zero $\mathbf{m}_{\mathbf{q}3k}$ (or any other higher order Fourier terms) increases $\bar{w}_{\mathrm{per}}$, its magnitude at equilibrium state is always at least an order of magnitude smaller than that of $\mathbf{m}_{\mathbf{q}1j}$.

**Method of soft-mode analysis for emergent crystals**. Based upon the qualitative analysis given above, we establish a method that tells us how to search for a possible solution, and then determine its metastability (i.e., if the solution corresponds to a local minimum of the free energy) by checking if there exist any soft modes[40] for the solution found. Through this method of soft-mode analysis, we obtain at the same time the possible directions of evolution if the solution is intrinsically unstable. Mathematically, any emergent crystalline state can be described by a characteristic vector $\mathbf{V}_{\mathrm{C}} = \begin{bmatrix} c_{111}^{\mathrm{re}} & c_{121}^{\mathrm{re}} & c_{131}^{\mathrm{re}} & c_{111}^{\mathrm{im}} & c_{121}^{\mathrm{im}} & c_{131}^{\mathrm{im}} \end{bmatrix}^T$ if its Fourier magnitudes monotonically decrease with the expansion order so that the first order Fourier components dominate its properties (e.g., the SkX can be described by $\mathbf{V}_{\mathrm{C}} = \frac{\sqrt{3}}{3}\begin{bmatrix}1 & 1 & 1 & 0 & 0 & 0\end{bmatrix}^T$). Nevertheless, we find that the value of $\mathbf{V}_{\mathrm{C}}$ is not unique for a certain type of emergent crystal. For example, SkX can be described by $\mathbf{V}_{\mathrm{C}1} = \frac{\sqrt{3}}{3}\begin{bmatrix}1 & 1 & 1 & 0 & 0 & 0\end{bmatrix}^T$ (Fig. 2e), or $\mathbf{V}_{\mathrm{C}2} = \begin{bmatrix}0.288 & -0.288 & -0.288 & 0.5 & 0.5 & 0.5\end{bmatrix}^T$ (Fig. 2d). This diversity derives from the arbitrariness of the choice of origin: the center of Wigner–Seitz cell can be displaced an arbitrary vector away from the center of a skyrmion in the SkX, which yields an infinite choice of $\mathbf{V}_{\mathrm{C}}$. On the other hand, DiskX can either be described by $\mathbf{V}_{\mathrm{C}a} = \frac{\sqrt{3}}{3}\begin{bmatrix}0 & 0 & 0 & 1 & 1 & 1\end{bmatrix}^T$ (Fig. 2a) or $\mathbf{V}_{\mathrm{C}b} = \frac{\sqrt{3}}{3}\begin{bmatrix}0 & 0 & 0 & -1 & -1 & -1\end{bmatrix}^T$ (Fig. 2b), since the two states (with field configuration opposite to each other) correspond to the same averaged free energy density in chiral magnets. As a result, the first step is to artificially choose an initial direction of $\mathbf{V}_{\mathrm{C}}$ (i.e., pre-assume that $\mathbf{V}_{\mathrm{C}0} = v_c\mathbf{U}_0$ where $\mathbf{U}_0$ is a known unit vector with six components and $v_c > 0$), and then solve $\frac{\partial \bar{w}(\mathbf{V}_{\mathrm{C}0})}{\partial m_{03}} = 0, \frac{\partial \bar{w}(\mathbf{V}_{\mathrm{C}0})}{\partial v_c} = 0, \frac{\partial \bar{w}(\mathbf{V}_{\mathrm{C}0})}{\partial q} = 0$ while setting all other variables not included in $\mathbf{V}_{\mathrm{D}}$ as zero. In the second step, numerically solve the free energy minimization problem of $\bar{w}$, selecting the solution obtained in the first step as the initial state, and denote the solution found by a vector $\mathbf{V}_0$ ($\mathbf{V}_0$ takes the same form as $\mathbf{V}$ introduced in Eq. (13)). In the third step, calculated the exact values of the matrix $\mathbf{D}$ by using $\mathbf{V}_0$, where the components of $\mathbf{D}$ are defined by $D_{ij} = \frac{\partial^2 \bar{w}}{\partial V_i \partial V_j}$, and $V_i$ denotes a component of the vector $\mathbf{V}$ introduced in Eq. (13). In the fourth step, solve the eigenvalue problem for the matrix $\mathbf{D}$ at $\mathbf{V}_0$. If all the eigenvalues are larger than zero, than $\mathbf{V}_0$ corresponds to an emergent crystalline state that is metastable. If there exist eigenvalues that are equal to or smaller than zero, then $\mathbf{V}_0$ corresponds to an emergent crystalline state that is intrinsically unstable. In this case, the $s$ negative eigenvalues denoted by $\lambda_1, \lambda_2, \ldots, \lambda_s$ characterize the relative magnitude of driving force of the soft modes, while the corresponding eigenvectors denoted by $\mathbf{R}_1, \mathbf{R}_2, \ldots, \mathbf{R}_s$ characterize the evolution direction of the soft modes. One can repeat all the steps introduced above by including the information of $\mathbf{R}_1, \mathbf{R}_2, \ldots, \mathbf{R}_s$ into the selection of the initial direction, in which case the characteristic vector $\mathbf{V}_C$ is replaced by the dominant vector $\mathbf{V}_D$ (i.e., $\mathbf{V}_{D1} = v_d\mathbf{U}_D + v_1\mathbf{V}_{R1} + v_2\mathbf{V}_{R2} + ... + v_s\mathbf{V}_{Rs}$, where $\mathbf{V}_{R1}$, $\mathbf{V}_{R2}$, ..., $\mathbf{V}_{Rs}$ are unit vectors extracted from $\mathbf{R}_1$, $\mathbf{R}_2$, ..., $\mathbf{R}_s$), and then do all the calculation again. By doing so, one derives the phase transition evolution path for an intrinsically unstable emergent crystal described by $\mathbf{V}_{C0}$, and then the destination of this evolution: another emergent crystalline state that is metastable. In practice, one will find that there always exist two eigenvalues of $\mathbf{D}$ very close to zero for any solution $\mathbf{V}_0$ which remain almost invariant with $b$. These two modes correspond to in-plane rigid body translation of the emergent crystals and should be distinguished from the real soft modes.

## Data availability

The data that support the plots within this paper and other findings of this study are available from the author on reasonable request.

## Acknowledgements

We thank Xiaoming Lan for verifying the validity and efficiency of the theory by performing comparison Monte Carlo simulations, and thank Prof. Bertrand Mercier for helpful discussion on functional analysis. The work was supported by the NSFC (National Natural Science Foundation of China) through the fund 11772360 and by the Pearl River Nova Program of Guangzhou through the fund 201806010134.

## Author contributions

Y.H. conceived the idea and conducted the work.

## Additional information

**Supplementary information** accompanies this paper at https://doi.org/10.1038/s42005-018-0071-y.

**Competing interests:** The author declares no competing interests.